# A novel space division multiplexing system for free space optical communications [*]

Zhou Hai-long (周海龙), Dong Jian-ji (董建绩) [+], Shi Lei (施雷), Huang De-xiu (黄德修), and Zhang Xin-Liang (张新亮)

Wuhan National Laboratory for Optoelectronics, School of Optoelectronic Science and Engineering, Huazhong University of Science and Technology, Wuhan 430074, China

We propose a novel space division multiplexing (SDM) technique based on spatial phase encoding, which may provide a new perspective beyond the conventional SDM in free space optical communications, such as multiplexing of orbital angular momentum (OAM). In our scheme, Gaussian beams are multiplexed with a set of spatial phase masks, and then transmitted in the Fourier domain via spherical lens. And another phase masks with matched patterns sandwiched by two lenses are used to de-multiplex the corresponding channels. Three different phase masks, namely, planar linear encoding, radial linear encoding and hybrid of radial and azimuthal linear encoding, are proposed and analyzed. The multiplexing and de-multiplexing of Gaussian beams are successfully implemented using these phase encoding approaches. We prove that the OAM multiplexing and radial phase encoding can be combined to further increase the communications capacity in free space.

## 1. Introduction

Space division multiplexing (SDM) in optical fiber link is expected to keep increasing the communication capacities besides the conventional multiplexing techniques [1, 2]. The fiber-based SDM techniques include mode-division multiplexing (MDM) using few-mode fibers (FMFs) [3-7], spatial multiplexing using multi-core fibers (MCFs) [8-11] or using helical excitation of different spatial angles [12], and orbital angular momentum (OAM) multiplexing using special ring fibers [13-16], etc. These new multiplexing techniques have also been introduced to free space optical communications. For example, free space OAM-based SDM has been attracting sustaining interests in recent years due to infinite channels of multiplexing [17-19]. Besides, the spatial phase encoding in free space is very easy to implement with binary optical elements, such as spatial light modulators, conical lenses, and blazed gratings, etc. However, people desire to exploit and develop new SDM approaches for further increasing the communication capacities.

[*] Project supported by the National Basic Research Program of China (Grant No. 2011CB301704), the Program for New Century Excellent Talents in Ministry of Education of China (Grant No. NCET-11-0168), a Foundation for the Author of National Excellent Doctoral Dissertation of China (Grant No. 201139), and the National Natural Science Foundation of China (Grant No. 60901006, and Grant No. 11174096).
[+] Corresponding author. Email: jjdong@mail.hust.edu.cn

In this paper, we propose a novel SDM system based on spatial phase encoding, in which three different phase encoding approaches are demonstrated. In our SDM system, spatial phase masks with different phase functions are used to encode the Gaussian beams for multiplexing. The multiplexed signals are transmitted in the Fourier domain via a spherical lens, and another phase mask with matched slope is used to de-multiplex the corresponding channel. The multiplexing and de-multiplexing of Gaussian beams are successfully implemented with three different phase masks, i.e., planar linear encoding, radial linear encoding, and hybrid of radial and azimuthal linear encoding. Our SDM system may provide a new perspective and technical reserve beyond the conventional SDM.

## 2. Beam energy transfer based on spatial phase encoding

De-multiplexing of SDM is to implement the spatial resolution of light beams at the receivers. The energy of light beams with appropriate spatial phase modulation can be transferred and identified via spatial Fourier transformation or via spatial diffraction in free space. Consequently, de-multiplexing is possible. For example, an OAM beam has an azimuthal phase distribution with $m(r,\theta) = \exp(-j\ell\theta)$, where $\ell$ is topological charge and $\theta$ is azimuthal angle. This phase distribution can be transformed to an annular beam at the Fourier diffraction plane [17]. Thus the beam energy is transferred from the center to the toroid. In this paper, we propose another two kinds of spatial phase modulation functions, $m(x,y) = \exp(jax)\exp(jby)$ and $m(x,y) = \exp(jar)$, where $x$ and $y$ are two-dimensional Cartesian coordinates, $r = \sqrt{x^2 + y^2}$, $j = \sqrt{-1}$, and $a, b$ are the modulation slopes. These two phase functions are named as planar linear phase encoding and radial phase encoding, respectively. Both functions can implement the beam energy transfer at the diffraction plane.

Figure 1 describes the schematic diagram of beam energy transfer with different spatial phase masks, where (a) shows a planar linear phase encoding with $\exp(jax)\exp(jby)$ and (b) shows a radial phase encoding with $\exp(jar)$. A spherical lens (Lens 1) is used to implement the spatial Fourier transform (SFT) from the front focal plane to the back focal plane. Assume that a spatial Gaussian beam is located at the center of the front focal plane, which may carry information of other possible dimensions, such as time, quadrature, frequency, and polarization, etc. A spatial phase mask then modulates the input Gaussian beam with a two-dimensional phase function of $m(x,y)$.

In Fig. 1(a), we define $m(x,y) = \exp(jax)\exp(jby)$. Then the wave propagation from the input-plane $(x_0, y_0)$ to the output-plane $(x_1, y_1)$ can be described as

$$E_{out}(x_1,y_1) \propto \mathscr{F}[E_{in}(x_0,y_0)m(x_0,y_0)] = V_{in}(\frac{x_1}{\lambda f} - \frac{a}{2\pi}, \frac{y_1}{\lambda f} - \frac{b}{2\pi}) \quad (1)$$

where $E_{in}$ is the input Gaussian beam, $E_{out}$ is the output field at the back focal plane of Lens 1. $\mathscr{F}[\cdot]$ denotes the spatial Fourier transform operation only on the transverse coordinates. $V_{in}(u,v)$ is the two-dimensional Fourier transform of $E_{in}(x_0, y_0)$, expressed by $V_{in}(u,v) = \mathscr{F}[E_{in}(x_0, y_0)]$. $\lambda$ is the wavelength in free space. $f$ is the lens focal length. From Eq. (1), one can see that the output beam is still a Gaussian beam but its coordinate shifts to $(af\lambda/2\pi, bf\lambda/2\pi)$, which is determined by the mask phase slope. Therefore, one can implement SDM of Gaussian beams using a set of phase masks with different phase slopes.

In Fig. 1(b) we define $m(x,y) = \exp(jar)$, i.e., radial phase encoding. Then the output beam can be expressed by

$$E_{out}(r_1,\theta_1) \propto \mathscr{F}[E_{in}(r_0,\theta_0)m(r_0,\theta_0)] = 2\pi\int_0^\infty r_0 E_{in}(r_0,\theta_0)m(r_0,\theta_0)J_0(2\pi r_0 \frac{r_1}{\lambda f})dr_0 \quad (2)$$

where $(r_0, \theta_0)$ and $(r_1, \theta_1)$ are the input-plane and output-plane in pole coordinates. And $J_0$ is the zero order Bessel function of the first kind. From Eq. (2), one can deduce by simulation that the output beam is an annular beam when $a \neq 0$. And the radius of the annular beam is determined by the mask phase slope. Therefore, one can implement SDM of Gaussian beams using radial phase encoding as well. From Figs. 1(a) and (b), one can see that the energy distribution of output beams will deviate from the axis center after modulation by the phase masks.

In terms of SDM de-multiplexing, we aim to convert these targeted beams whose energy distribution is off the axis center to be on the axis center, while other untargeted channels keep the energy distribution off the axis center. This is the basic design idea for our SDM system, which will be discussed in detail in the next section.

## 3. Implementation of our SDM

Based on the aforementioned principle, we design a novel SDM system to achieve the multiplexing and de-multiplexing of information-carrying Gaussian beams in free space, as shown in Fig. 2. Define that $U_{in1}$, $U_{in2}$ and $U_{in3}$ are the optical field at the front focal plane of Lens 1, Lens 2 and Lens 3 respectively, and $U_{out1}$, $U_{out2}$ and $U_{out3}$ are the optical field in the back focal plane of Lens 1, Lens 2 and Lens 3 respectively. Assume that $E_{in,l}(l = 1,2 \ldots n)$ are input Gaussian beams carrying corresponding data. Subscript $l$ represents different data channels to be multiplexed, and subscript $in$ represents the input spatial Gaussian beam. All these channels are modulated by different phase masks, whose transfer functions are labeled by $m_{in,l}$. Then all these channels are multiplexed to one channel for transmission. Subsequently the multiplexed signals will be converted into the Fourier domain through Lens 1, and then propagate in free space for a distance of $z$, the transfer function of Fresnel diffraction for a distance of $z$ can be expressed as

$$H(\mu,\nu) = \exp(jkz)\exp[-j\pi\lambda z(\mu^2 + \nu^2)] \quad (3)$$

where $k$ is the wavenumber. Define $H_{id}(\mu,\nu) = \mathscr{F}[U_{id}(x,y)]$, where subscript $id = in1$, $in2$, $in3$, $out1$, $out2$, $out3$. Another phase mask with a phase pattern $m_2(x,y)$ sandwiched by two lenses (Lens 2 and Lens 3) is used to de-multiplex the corresponding channels. Define $(x_0, y_0), (x_2, y_2), (x_4, y_4)$ as the front focal plane of Lens 1, Lens 2 and Lens 3 respectively, and $(x_1, y_1), (x_3, y_3), (x_5, y_5)$ as the back focal plane of Lens 1, Lens 2 and Lens 3 respectively. Take $E_{in,l}$ as an example, it can be easily derived that

$$U_{in1}(x_0, y_0) = E_{in,l}(x_0, y_0)\,\mathrm{m}_{in,l}(x_0, y_0) \quad (4)$$

$$U_{out1}(x_1, y_1) = \frac{\exp(2jkf)}{j\lambda f} H_{in1}(\frac{x_1}{\lambda f}, \frac{y_1}{\lambda f}) \quad (5)$$

$$H_{in2}(\mu,\nu) = H_{out1}(\mu,\nu)H(\mu,\nu) \quad (6)$$

$$U_{out2}(x_3, y_3) = \frac{\exp(2jkf)}{j\lambda f} H_{in2}(\frac{x_3}{\lambda f}, \frac{y_3}{\lambda f}) \tag{7}$$

$$U_{in3}(x_4, y_4) = U_{out2}(x_4, y_4) m_2(x_4, y_4) \tag{8}$$

Substitute Equations (3) - (7) to Eq. (8), we can get

$$U_{in3}(x_4, y_4) = -E_{in,l}(-x_4, -y_4) m_{in,l}(-x_4, -y_4) m_2(x_4, y_4)$$
$$\cdot \exp(4jkf)\exp(jkz)\exp[-j\frac{\lambda z\pi}{(\lambda f)^2}(x_4^2 + y_4^2)] \tag{9}$$

In Eq. (9), $\exp(4jkf)\exp(jkz)$ is a complex constant, and the quadratic phase function $\exp[-j\lambda z\pi(x_4^2 + y_4^2)/(\lambda f)^2]$ can also be regarded as a constant if the propagation distance in free space $z$ is not very long or the waist size of input Gaussian beam is not very large. Then Eq. (9) can be simplified by

$$U_{in3}(x_4, y_4) = \gamma E_{in,l}(-x_4, -y_4) m_{in,l}(-x_4, -y_4) m_2(x_4, y_4) \tag{10}$$

where $\gamma$ is a constant. Lens 3 is used to implement SFT from $U_{in3}$ to $U_{out3}$. From Eq. (10), one can see that the output beam $U_{out3}$ will turn to be a Gaussian beam with the energy profile at the axis center if and only if $m_{in,l}(-x, -y)m_2(x, y)$ is a constant. Otherwise, the output beam will be off the center. In this way, the de-multiplexing of all channels can be obtained by designing the phase mask. For example, if the multiplexing phase mask is designed by $m_{in,l}(x, y) = \exp(jax)\exp(jby)$, then the de-multiplexing phase mask should be $m_2(x, y) = \exp(jax)\exp(jby)$, and if we design $m_{in,l}(x, y) = \exp(jar)$, then $m_2(x, y) = \exp(-jar)$. Besides, if we design $m_{in,l}(x, y) = \exp(j\ell\theta)$, then $m_2(x, y) = \exp(-j\ell\theta)$, where $\ell$ is topological charge. This case turns to be the multiplexing of OAM beams. Therefore, our SDM structure is suitable for OAM multiplexing as well. In the following, the first two spatial phase masks will be discussed. First, the planar linear phase encoding is analyzed. As an example, assume that $f = 100mm$, $\lambda = 1.55um$, $z = 1m$, $\omega = 0.1mm$, where $\omega$ is the waist size of the input Gaussian beams. Define that four dependent Gaussian beams are phase modulated by phase masks of $m_{in,l}(x, y) = \exp(jn_x ax)\exp(jn_y ay)$, where $n_x, n_y \in \{-0.5, 0.5\}$, $a = 100mm^{-1}$. Subscript $l$ represents individual channel. Subsequently these four beams are multiplexed to one channel, whose superposed energy is shown in the upleft of Fig. 3 (a). An interference pattern is seen due to the superposition of phase information. $U_{in1}$ is transformed to the output optical field $U_{out1}$ using the SFT of Lens 1, showing four separate Gaussian beams. These four Gaussian beams are tagged from number 1 to 4.

$U_{out1}$ is then transmitted for a distance of 1 meter in free space, whose optical field is shown as $U_{in2}$. We can see that the waist size of these beams is expanded but the spatial position is invariable, $U_{in2}$ is then converted to $U_{out2}$ with the SFT again by Lens 2. These separate beams are converted to one much smaller beam with an interference pattern, as shown in the downright of Fig. 3(a). The optical field of $U_{out2}$ is then modulated by the second matched phase mask $m_2(x, y)$, which is used to retrieve the Gaussian beam of No. 1. Then the output beam of $U_{out3}$ is calculated as shown in the upleft of Fig. 3(b). We notice that all the separate Gaussian beams are off the center except Beam 1. If a spatial filter is located at the central axis to distinguish Beam 1, then the first channel could be successfully de-multiplexed.

In order to de-multiplex all the four channels, a set of matched phase masks are employed. Figure 3(b) shows the de-multiplexed energy profiles of all channels. Only the channel of interest is decoded to the center by its matched phase mask, and the other channels are decoded to the off-center distribution. This proves the feasibility of our SDM system. Besides, it should be noted that the energy pattern of $U_{out3}$ is rotated by 180 degree compared to $U_{in2}$, which can be explained by Eq. (11).

$$U_{out3}(x_4, y_4) \propto \mathscr{F}\left\{H_{in2}(\frac{x_3}{\lambda f}, \frac{y_3}{\lambda f})\exp(jax_3)\exp(jby_3)\right\}_{\mu=\frac{x_4}{\lambda f}, \nu=\frac{y_4}{\lambda f}} \quad (11)$$
$$= U_{in2}(-x_4 + \frac{af\lambda}{2\pi}, -y_4 + \frac{bf\lambda}{2\pi})$$

Free-space propagation of Gaussian beams can be considered as a spatial diffusion process. And the waist size of the Gaussian beams will be expanded along the propagation distance $z$. To avoid the channel crosstalk, the overlapped intensity profile of two adjacent beams should be as small as possible, which is defined to be smaller than 50% of the maximum light spot energy in our analysis. For the planar linear phase encoding, assume that the phase masks are expressed as $m_{in,l}(x, y) = \exp(ja_l x) \exp(jb_l y)$, and the step of adjacent phase slopes in x-axis and y-axis are both $da$. From Eq. (9), we can derive the maximum transmission distance $z_{max}$ via Fourier transform.

$$z_{max} = \frac{f^2 \lambda}{\pi} \sqrt{\frac{da^2}{4A_{th}^2 \omega^2} - \frac{1}{\omega^4}} \quad (12)$$

where $A_{th}$ is the minimum distance between two adjacent spots over the waist size of output beams, and $\omega$ is waist size of input Gaussian beams. It can be calculated that $A_{th} = 2.04$ for Gaussian beams. From Eq. (12), we can find that the maximum transmission distance have a threshold condition expressed as $da > 2A_{th}/\omega$. And only if the threshold condition is satisfied, the beams can be transmitted with low crosstalk.

Figure 4 shows the maximum transmission distance as a function of phase slope step and waist size respectively when $f = 100mm$. The maximum transmission distance can be extended as the phase slope step increases. The reason lies in that a large phase slope can make the distance of two adjacent beams larger, resulting in a low crosstalk. Fig. 4(b) reveals that the maximum transmission distance has an optimum, which is determined by $\omega = 2\sqrt{2}A_{th}/da$. The maximum transmission distance decreases as the waist size of Gaussian beam keeps further increasing.

In the following, the radial phase encoding is analyzed. Assume that $f = 100mm, \lambda = 1.55um, \omega = 0.1mm$ and $z = 1m, a = 100mm^{-1}, m_{in,l}(x,y) = \exp(jn_r ar)$, where $n_r \in \{-1,0,1\}$. Subscript $l$ represents three individual channels. Fig. 5 shows the simulated energy profiles at different planes. Fig. 5 (a) shows the propagation of the three beams modulated by $\exp(jn_r ar)$ where $n_r = -1,0,1$ respectively. Similar to Fig. 3 (a), $U_{in1}$ is transformed to $U_{out1}$ using the SFT of Lens 1. It is proved that an annular beam is generated from a Gaussian beam when $n_r = -1, 1$. $U_{out1}$ is then transmitted for a distance of 1 meter in free space, whose optical field is shown as $U_{in2}$. One can see that the beams size is expanded when $n_r = -1,0$ and shrunk when $n_r = 1$. The reason lies in that the diffraction-induced spatial dispersion is accumulated when $n_r = -1,0$ but compensated when $n_r = 1$. $U_{in2}$ is then converted to $U_{out2}$ with the SFT again by Lens 2. The energy profiles for all channels return to be light spots. Fig. 5 (b) shows the de-multiplexing of three beams separately, where a1, b1, and c1 are output energy profiles of three beams when $m_2(x,y) = \exp(jar)$, a2, b2, and c2 are output energy profiles when $m_2(x,y) = 1$, a3, b3, and c3 reveal output energy profiles when $m_2(x,y) = \exp(-jar)$. It can be seen that only the channel of interest is decoded to the axis center by its matched phase mask, such as a1, b2, c3 of Fig. 5(b). At the same time, all the other channels are decoded to annular beams with off-axis distribution.

It should be noted that energy distributions in Fig. 5 are normalized so that they can be clear seen. In fact, the peak energy of an annular beam is much lower than the peak of a Gaussian beam after beam transformation. This is very helpful to distinguish the central

Gaussian beam from the surround annular beams with low channel crosstalk. To analyze the crosstalk of different channels, a crosstalk coefficient $\eta$ is defined as the energy integral of all the other channels over the energy integral of the channel of interest along a certain plane. Ignoring the crosstalk of high order channels, $\eta$ can be expressed by

$$\eta = \frac{\iint\limits_{r<r_{HM}} (I_1 + I_{-1}) dxdy}{\iint\limits_{r<r_{HM}} I_0 dxdy} . \tag{13}$$

where $I_0$ is the output energy profile of channel of interest, such as b2 of Fig. 5(b), $I_1$ and $I_{-1}$ are the two adjacent energy profiles out of interest, such as a2 and c2 of Fig. 5(b), $r_{HM}$ is the radius at the half of $I_0(0, 0)$. The maximum transmission distance $z_{max}$ is restrained by a certain crosstalk coefficient $\eta$. For instance, Fig. 6 shows the simulated results of the maximum transmission distance as a function of the lens focal length when $\eta = 10\%$ and $\eta = 1\%$. In the simulations, define the step of adjacent radial phase slopes is $da$. From Fig. 6, we find that $z_{max}$ is proportional to the square of lens focal length and have a threshold condition as well. And $z_{max}$ also positively correlates to the radial phase step. Similar to Eq. (12), we can obtain the estimate equation of $z_{max}$ by curve-fitting method, as shown in Eq. (14). For $\eta = 10\%$, we estimate $A_{th}$=2.143 and $K$=0.3687. For $\eta = 1\%$, we estimate $A_{th}$=3.225, and $K$=0.5186.

$$Z_{max} = Kf^2\lambda\sqrt{\frac{da^2}{4A_{th}^2\omega^2} - \frac{1}{\omega^4}} . \tag{14}$$

We have proposed two phase masks, i.e., linear phase encoding and radial phase encoding, to implement SDM in free space. The performances of these SDM approaches can be compared with OAM multiplexing, as shown in Table 1. The phase masks in our schemes may be much simpler than helical phase mask in OAM system. The multiplexing/de-multiplexing of our schemes is related to the slope of phase masks, while OAM channels are distinguished by topological charge. All these SDM schemes can support an infinite channel capacity, but support a short transmission distance in free space due to the limitation of spatial diffraction. It was reported that ring fibers could support OAM modes, but the transmission distance was still very short as well. Besides, the radial phase mask may induce a lower crosstalk compared to the linear phase mask, since the beams of irrelative channels are transformed to annular beams with very low energy density, while the relative channel is transformed to the axis center with a high energy.

Table 1. SDM comparison in free space

| SDM | Linear phase encoding | Radial phase encoding | Azimuthal phase encoding (OAM) |
|---|---|---|---|
| Phase mask | $\exp(jax)\exp(jby)$ | $\exp(jar)$ | $\exp(j\ell\theta)$ |
| Multiplexing parameter | phase slope $(a, b)$ | phase slope $(a)$ | topological charge $(\ell)$ |
| Propagating distance | short | short | short |
| Channel capacity | infinite | infinite | infinite |
| Beam shape | Gaussian beam | annular beam | annular beam |
| Available in fiber | Not yet | Not yet | Yes, but short distance |

Figures 7 (a) and (b) show the energy transfer of Gaussian beams using azimuthal phase mask and radial phase mask, respectively. Since both radial phase mask and azimuthal phase mask of OAM could result in an annular beam, the hybrid multiplexing of these two phase masks is possible. Fig. 7 (c) shows the principle of the hybrid multiplexing of OAM and radial phase encoding and the output beam is still an annular beam. Assume that we have four channels which are modulated by $m_{in,l}(x,y) = exp(ja_l r)exp(j\ell_l\theta)$ where $(a_l, \ell_l) = (-50mm^{-1}, -1), (50mm^{-1}, -1), (-50mm^{-1}, 1)$, and $(50mm^{-1}, 1)$ respectively. And define $f = 100mm, \lambda = 1.55um, z = 1m, \omega = 0.1mm$. Fig. 8 shows the de-multiplexing of these four channels, where the four rows display the output intensity ($U_{out3}$) of these channels under different decoding phase masks $m_2(x,y) = exp(jar)exp(j\ell\theta)$ respectively. From Column 1 to Column 4, the parameter pair $(a, \ell)$ equals to $(50mm^{-1}, 1)$ $(-50mm^{-1}, 1)$, $(50mm^{-1}, -1)$, and $(-50mm^{-1}, -1)$, respectively. It can be seen that only the channel of interest is decoded to the axis center by its matched phase mask, such as the patterns in diagonal line of Fig. 8. At the same time, all other channels are decoded to annular beams with off-axis distribution. It proves that the OAM multiplexing and radial phase encoding can be combined to further increase the communications capacity.

## 4. Conclusions

We have proposed a novel SDM technique, which may provide a new perspective beyond the conventional SDM in free space optical communications. Our SDM technique is based on spatial phase slope encoding of Gaussian beams. The multiplexed signals are transmitted in the Fourier domain with SFT of spherical lens, and a matched phase mask sandwiched by two spherical lenses is used to de-multiplexing. The multiplexing and de-multiplexing of Gaussian beams are successfully implemented with three different phase masks, i.e., planar linear encoding, radial linear encoding and hybrid of radial and azimuthal linear encoding. The optimization of our SDM scheme is also discussed in terms of the maximum propagation distance. We also prove that the OAM multiplexing and radial phase encoding can be combined to further increase the communication capacity.

Figure Captions:

**Fig. 1.** Beam energy transfer with different spatial phase masks, (a) planar linear phase encoding with $\exp(jax)\exp(jby)$, and (b) radial phase encoding with $\exp(jar)$.

**Fig. 2.** Schematic diagram of the proposed SDM using spatial phase slope encoding.

**Fig. 3.** Simulation results of multiplexing/demultiplexing of four spatial phase modulation beams. (a) the propagation of the four beams modulated by $\exp(jn_x ax)\exp(jn_y ay)$, $n_x, n_y \in \{-0.5, 0.5\}, a = 100mm^{-1}$, (b) de-multiplexed beams of No. 1-4, respectively.

**Fig. 4.** The maximum transmission distance under different step $da$ and waist size $\omega$ for the first phase mask $\exp(jax)\exp(jby)$ when $f = 100mm$.

**Fig. 5.** Simulation results of multiplexing/demultiplexing of three spatial phase modulation beams. (a) the propagation of the three beams modulated by $m_{in,l}(x,y) = \exp(jn_r ar), n_r \in \{-1,0,1\}, a = 100mm^{-1}$. (b) de-multiplexing of the three beams.

**Fig. 6.** simulation result of the maximum transmission distance $z_{max}$ when $\eta = 10\%$ and $\eta = 1\%$ ( $\omega$ contribution in *mm*, $da$ contribution in *mm*$^{-1}$)

**Fig. 7.** the principle of this kind hybrid multiplexing of radial phase mask and azimuthal phase mask of OAM. (a) azimuthal phase of OAM. (b) radial phase. (c) hybrid phase

**Fig. 8.** de-multiplexing of the four beams modulated by $m_{in,l}(x,y) = exp(ja_l r)exp(j\ell_l \theta)$ where $(a_l, \ell_l) = (-50mm^{-1}, -1), (50mm^{-1}, -1)(-50mm^{-1}, 1), (50mm^{-1}, 1)$ respectively.

Figures:

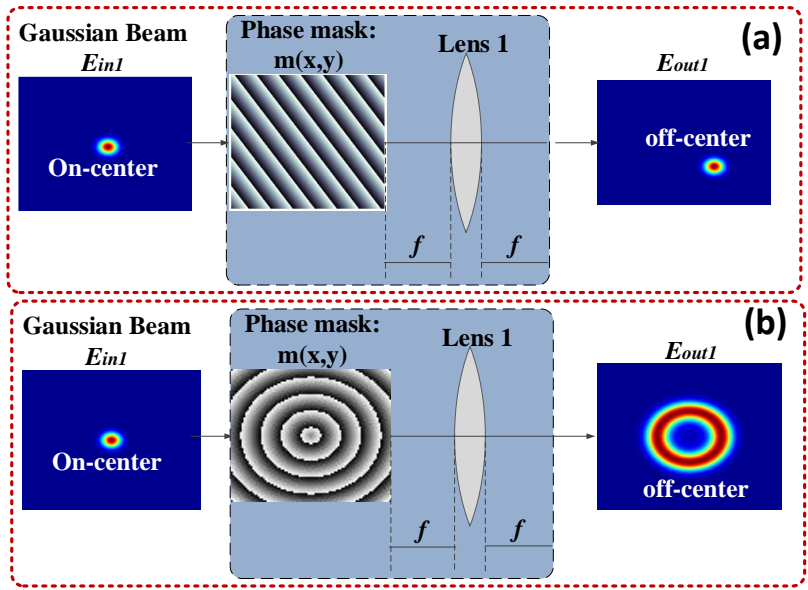

**Fig. 1**

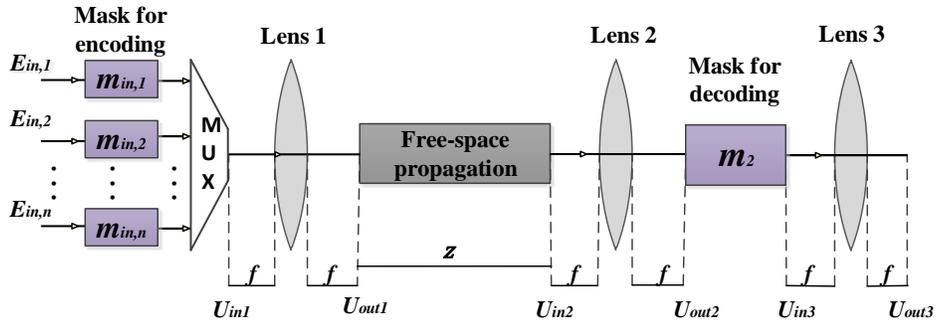

**Fig. 2**

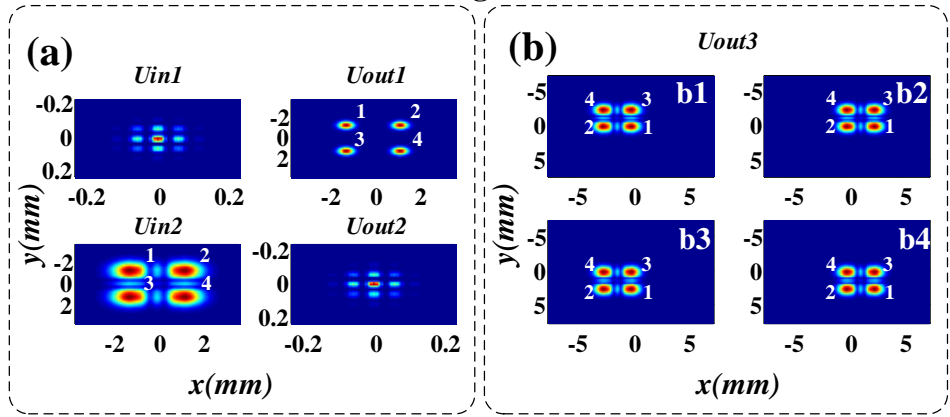

**Fig.3**

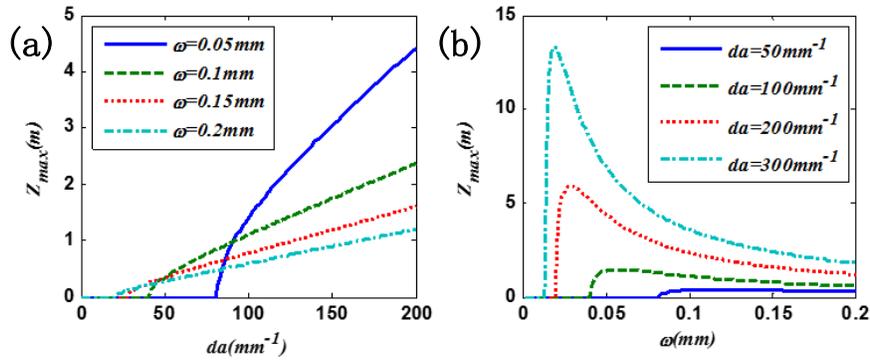

**Fig.4**

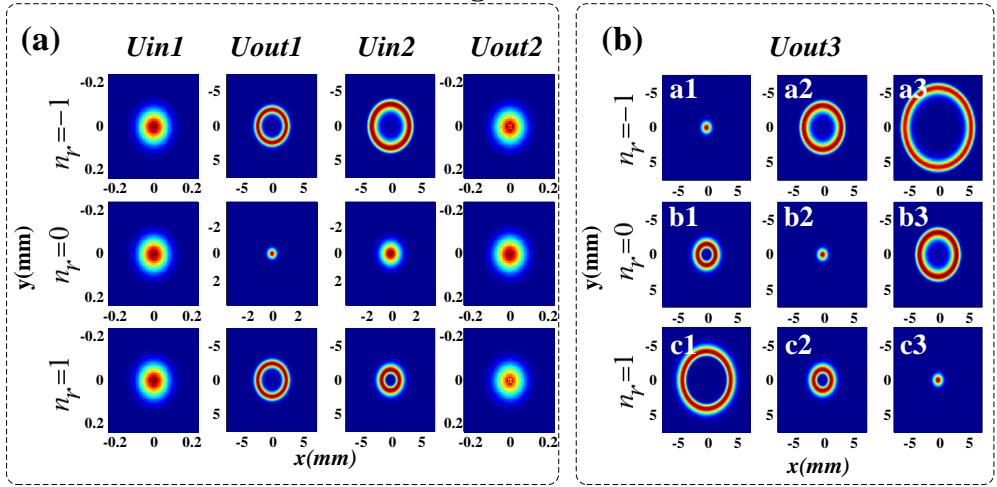

**Fig.5**

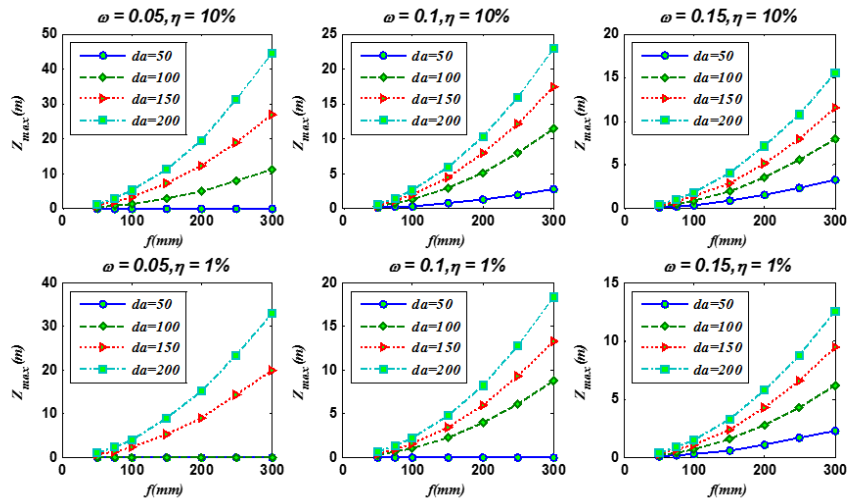

**Fig.6**

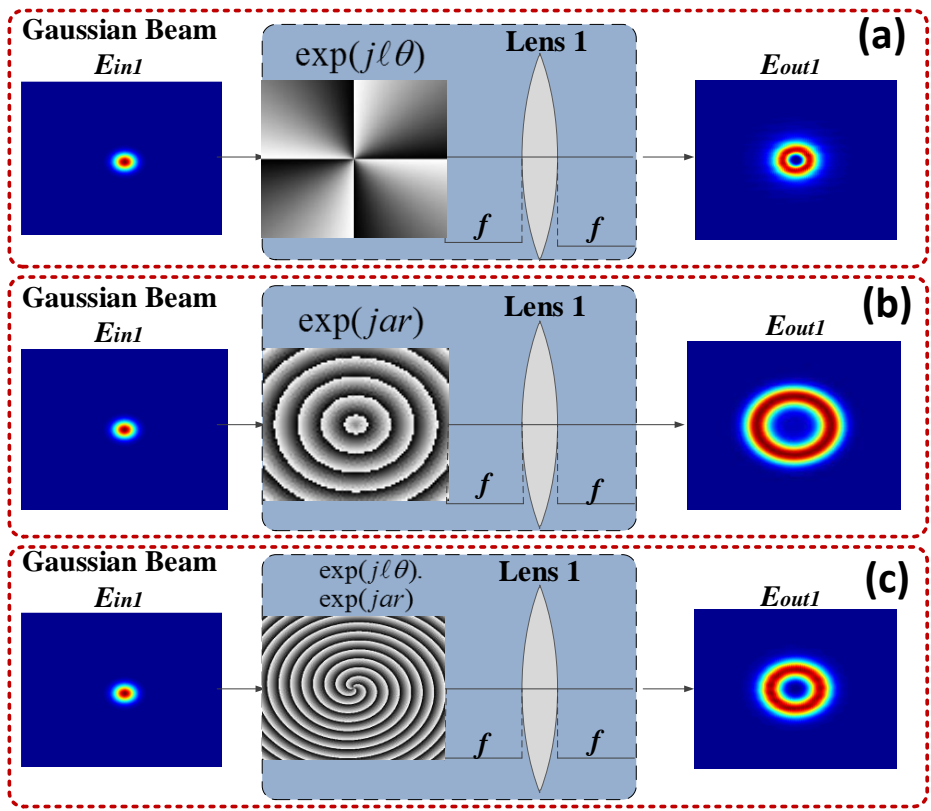

**Fig. 7**

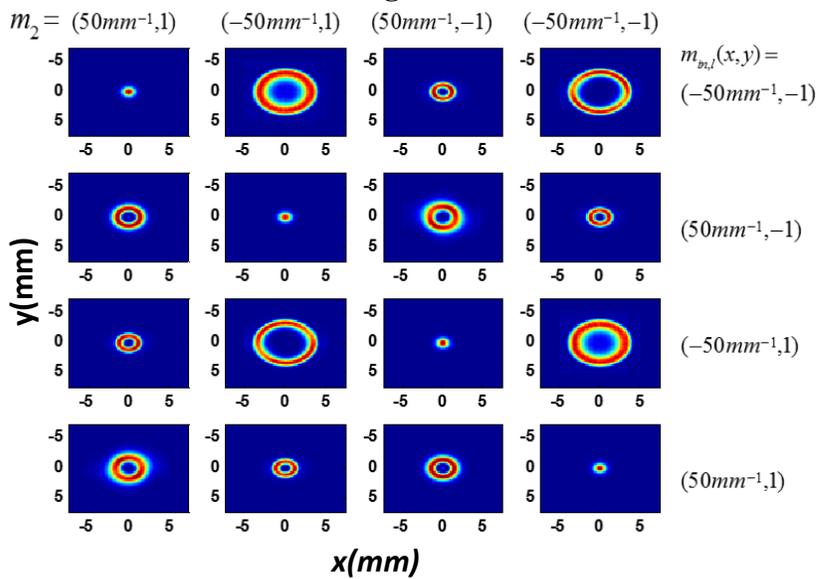

**Fig. 8**